# Scalable production of single 2D van der Waals layers through atomic layer deposition: Bilayer silica on metal foils and films

Gregory S. Hutchings[1], Xin Shen[1], Chao Zhou[2], Petr Dementyev[3], Daniil Naberezhnyi[3], Inga Ennen[3], Andreas Hütten[3], Nassar Doudin[1], Jesse Hsu[1], Zachary S. Fishman[1], Udo D. Schwarz[1,2], Shu Hu[1] and Eric I. Altman[1]

[1] Department of Chemical and Environmental Engineering, Yale University, New Haven, CT 06520, USA

[2] Department of Mechanical Engineering and Materials Science, Yale University, New Haven, CT 06520, USA

[3] Faculty of Physics, Bielefeld University, 33615 Bielefeld, Germany

## Abstract

The self-limiting nature of atomic layer deposition (ALD) makes it an appealing option for growing single layers of two-dimensional van der Waals (2D-VDW) materials. In this paper it is demonstrated that a single layer of a 2D-VDW form of $SiO_2$ can be grown by ALD on Au and Pd polycrystalline foils and epitaxial films. The silica was deposited by two cycles of bis (diethylamino) silane and oxygen plasma exposure at 525 K. Initial deposition produced a three-dimensionally disordered silica layer; however, subsequent annealing above 950 K drove a structural rearrangement resulting in 2D-VDW; this annealing could be performed at ambient pressure. Surface spectra recorded after annealing indicated that the two ALD cycles yielded close to the silica coverage obtained for 2D-VDW silica prepared by precision SiO deposition in ultra-high vacuum. Analysis of ALD-grown 2D-VDW silica on a Pd(111) film revealed the co-existence of amorphous and incommensurate crystalline 2D phases. In contrast, ALD growth on Au(111) films produced predominantly the amorphous phase while SiO deposition in UHV led to only the crystalline phase, suggesting that the choice of Si source can enable phase control.

## Introduction

Two-dimensional van der Waals (2D-VDW) materials have attracted a great deal of attention over the last decade because of their stability as individual atomic layers, the unique properties that can emerge when they are reduced to their atomic limit, the potential to tease out novel properties by stacking complementary layers, and the ability to fabricate devices by simple mechanical stacking [1-4]. While progress has been made by thinning bulk 2D materials, it was realized early on that practical applications require controllable growth of one to several 2D-VDW layers [5, 6]. Growth also opens the possibility of generating 2D-VDW layers that typically do not exist in bulk layered forms. Thus, a range of growth processes based on chemical vapor deposition, vapor transport, and reaction and surface segregation have been developed [7-12]. Challenges in reproducibly fabricating single 2D-VDW layers with large domains include growth rates that depend strongly on not only the substrate composition but also its crystallographic orientation, defect density, and thickness, which lead to sensitivities to the reaction or deposition time and temperature [6, 13-15]. On the surface, atomic layer deposition (ALD) would appear to be an appealing way to controllably produce any number of 2D-VDW layers on demand. Ideally, the self-limiting nature of each ALD half-cycle would produce a single 2D-VDW layer; in practice, however, the situation is much more complex. Typically, many cycles are required to create a single layer, and given the intrinsic inertness of 2D-VDW basal planes, even more cycles are necessary to produce additional layers [16, 17]. Further complications include disparities in the temperature requirements for the self-limiting reactions and crystallization [18]. In this paper, it will be shown that these challenges can be overcome for 2D-VDW $SiO_2$, a unique polymorph of silica that can be a structural template for other 2D tetrahedral oxides, and contains the building block of a wide range of 2D silicates [12, 19-31].

Two-dimensional van der Waals $SiO_2$ is constructed of two mirror image planes of rings of corner-sharing $SiO_4$ tetrahedra [19-25]. It can adopt a crystalline form with only six-membered rings and an amorphous form with four- through nine-membered rings [19-25]. Heretofore 2D-VDW $SiO_2$ has been grown on late transition metal surfaces (Ru, Pd, Pt, and Ni-Pd alloys) by ultra-high vacuum (UHV) molecular beam epitaxy (MBE) of just the right amount of either Si or SiO at mild temperatures (<525 K) followed by annealing above 975 K in low pressure oxygen (~$10^{-6}$ Torr) [23-25, 32, 33]. Two-dimensional $SiO_2$ has also been observed on graphene on Cu but the growth in this case was ill-defined [21]. The substrate, annealing time and temperature, silica coverage, and cooling rate appear to all play roles in the competition between the crystalline and amorphous forms, but thus far selection between the two phases has proven difficult [34, 35].

Unlike other 2D-VDW materials, the 2D silica bilayer structure intrinsically includes small molecule-sized (e.g., water) openings making it attractive as an ultimate permeation membrane [36, 37]. When doped with Al it displays similar local chemistry to zeolites, making it of interest in catalysis [28, 38]. As a small molecule-permeable 2D-VDW layer it has potential in the emerging area of catalysis under confinement [39-41]. Other potential applications include dielectric layers in VDW heterostructures. It has recently been shown that $GeO_2$ can form a similar 2D-VDW amorphous phase and a theoretical screening highlighted the potential for $AlPO_4$ and $GaPO_4$ to adopt the crystalline 2D-VDW bilayer structure [26, 34]. Therefore, 2D-VDW $SiO_2$ can be considered a prototype for tetrahedrally-coordinated 2D materials. It has also been shown that 2D-VDW silica can be released from its growth substrate both by delamination and by etching [21, 42]. Thus, a key missing piece in furthering research into and application of 2D-VDW silica and related materials is a simple, precise, reproducible, conformable, and scalable non-UHV growth method. The ALD method introduced here meets these criteria, as the few-cycle ALD process with precursors dispensed in a spatial sequence is compatible with roll-to-roll manufacturing.

**Results and Discussion**

The target substrates chosen were polycrystalline Pd, Au, and Cu foils and epitaxial Pd(111) and Au(111) thin films. Palladium was chosen based on prior successful 2D-VDW $SiO_2$ growth on Pd(100) and (111) [24, 43, 44], Au because its inertness enables processing in air, and Cu because of its low cost. The foils were selected for their potential in roll-to-roll processes. Meanwhile, the epitaxial Pd(111) film offered the opportunity for detailed structural characterization by low energy electron diffraction (LEED) and scanning tunneling microscopy (STM), and epitaxial Au(111) access to etching methods that allow characterization by high resolution transmission electron microscopy (HR-TEM).

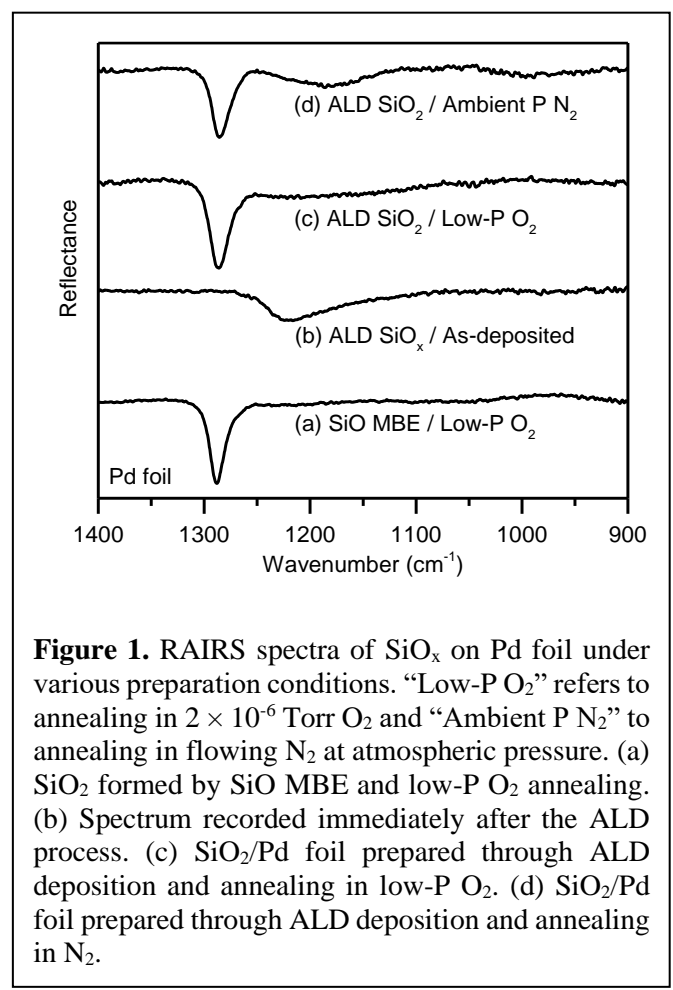

**Figure 1.** RAIRS spectra of $SiO_x$ on Pd foil under various preparation conditions. "Low-P $O_2$" refers to annealing in $2 \times 10^{-6}$ Torr $O_2$ and "Ambient P $N_2$" to annealing in flowing $N_2$ at atmospheric pressure. (a) $SiO_2$ formed by SiO MBE and low-P $O_2$ annealing. (b) Spectrum recorded immediately after the ALD process. (c) $SiO_2$/Pd foil prepared through ALD deposition and annealing in low-P $O_2$. (d) $SiO_2$/Pd foil prepared through ALD deposition and annealing in $N_2$.

As shown in previous studies, reflection-absorption infrared spectroscopy (RAIRS) can be used to distinguish two-dimensional phases of silicates on metal surfaces, with the benefits that neither atomic-level flatness nor metal facet control are necessary [23]. In addition, the inertness of 2D-VDW $SiO_2$ allows the measurements to be performed at ambient pressure. These benefits make the technique ideal for primary identification of silicates on metal foils. To provide a benchmark, 2D-VDW $SiO_2$ was grown on a Pd foil using SiO MBE. As shown in Figure 1a, the MBE-grown sample displays a distinct vibration at 1289 cm$^{-1}$, which is consistent with previous reports for 2D-VDW $SiO_2$ on Pd(111) and matches the frequency expected for stretching the 180° Si-O-Si bond that joins the two halves of the bilayer [23, 33]. Notably there are no features near 1000 cm$^{-1}$ that are indicative of Si-O-metal linkages [23, 29]. In a parallel experiment, controlled overdeposition to 2.3 ML (where 1 ML = $8.2 \times 10^{14}$ Si/cm$^{-2}$, half the amount of Si to form the 2D-VDW bilayer structure) through MBE results in a second RAIRS peak at 1265 cm$^{-1}$ (see Supporting Information). Peaks near 1260 cm$^{-1}$ have previously been reported following high temperature annealing of 4 ML $SiO_2$/Ru(0001), and several nm thick silica films grown on Mo and have been assigned to vibrational modes in quartz-like structures [23]. Thus, RAIRS can be used to sensitively distinguish if the target silica coverage is exceeded.

Atomic layer deposition was performed using cycles consisting of exposure to the standard ALD silica precursor bis(diethylamino)silane and an oxygen plasma,[45] both with the sample temperature fixed at 525 K. The substrates were pre-exposed to the oxygen plasma prior to the ALD cycles to remove any carbonaceous deposits. Initial experiments were performed on substrates that were "pre-cleaned" by UHV sputter-annealing cycles before transfer through air to the ALD system. Two ALD cycles were tested which follows the required silica coverage based on the 2D-VDW silica bilayer structure. As illustrated in the RAIRS spectrum in **Figure 1b**, the as-deposited silica exhibits only a broad feature centered around 1220 cm$^{-1}$, which can be attributed to a disordered 3D silica layer.

Annealing this layer above 975 K in $2\times10^{-6}$ Torr $O_2$ (**Figure 1c**) led to a sharp peak at 1286 cm$^{-1}$, consistent with reorganization of the silica into the 2D-VDW structure. Because the ALD system cannot reach the required annealing temperature, the annealing was performed after sample transfer to a UHV system. Insufficient annealing close to the threshold for bilayer formation resulted in coexistence of this broad feature and the expected bilayer $SiO_2$ mode. The slight difference in the IR peak position for the MBE and ALD grown materials can be due to differences in adsorption at the 2D-VDW metal interface [46]. In particular, water and oxygen have been found to preferentially adsorb at the 2D silica/Pd interface [39, 47]. While no impurities on the Pd sample were detected after re-loading the sample into the UHV system and anealing, AES cannot distinguish adsorbed oxygen, oxygen in water, and oxygen that is part of the 2D material [39, 40, 46].

The data described above were for Pd foils pre-cleaned by sputter/anneal cycles in a UHV system and returned to the same UHV system for the final annealing step. To test whether these two demanding UHV processing steps were needed, two additional samples were prepared by ALD: 1) no UHV pre-clean followed by low pressure oxygen anneal in the UHV system; and 2) UHV pre-clean followed by annealing in a furnace in flowing $N_2$ after ALD. Nitrogen was used rather than air to keep the oxygen partial pressure low enough to avoid Pd oxide formation. At atmospheric pressure just trace amounts of $O_2$ are required to avoid $SiO_2$ decomposition, an $O_2$ partial pressure of $4 \times 10^{-8}$ Torr that corresponds to less than 0.1 ppb [27]. The RAIRS spectra for both samples showed peaks at the same 1286 cm$^{-1}$ as the sample that was pre- and post-deposition processed in UHV; this is shown in **Figure 1d** for the ambient pressure $N_2$-annealed sample. Thus, neither UHV processing step is necessary. While $N_2$ is mostly inert, X-ray photoelectron spectroscopy (XPS) indicated a faint trace of residual N. The N is not due to the N in the precursor; N was below the XPS detection level for layers annealed in the UHV system (see Supporting Information). The N incorporation can be avoided by annealing in Ar instead of $N_2$.

To further evaluate the structure of ALD-deposited 2D-VDW $SiO_2$, ALD was used to deposit silica onto an epitaxial Pd(111) thin film. This silica layer was grown using two ALD cycles followed by annealing in low pressure $O_2$ in the UHV system; the RAIRS spectrum showed a peak near 1290 cm$^{-1}$. The LEED pattern for this sample, **Figure 2a**, exposes a ring of continuous intensity with 12 broad spots superimposed. Analysis of the spot positions in LEED patterns collected at different energies (see Supporting Information) reproduce prior work on the formation of incommensurate crystalline 2D-VDW SiO$_2$ on Pd(111) in UHV by SiO deposition [43]. Meanwhile, the continuous faint ring is typically assigned to amorphous 2D-VDW SiO$_2$ [23, 42, 44, 48], which has also been reported on Pd(111) [44]. The STM images in **Figure 2b,c** bear out these assignments. The image in **Figure 2b** reveals the hexagonal structure of a crystalline domain while the image in **Figure 2c** displays a mostly amorphous region with pockets of crystalline order extending a few unit cells. The coexistence of the amorphous and incommensurate crystalline phases in the ALD grown material is distinct from the prior work, which indicated the formation of the amorphous phase when elemental Si deposition was the starting point and the crystalline phase when the growth started with SiO deposition [43, 44].

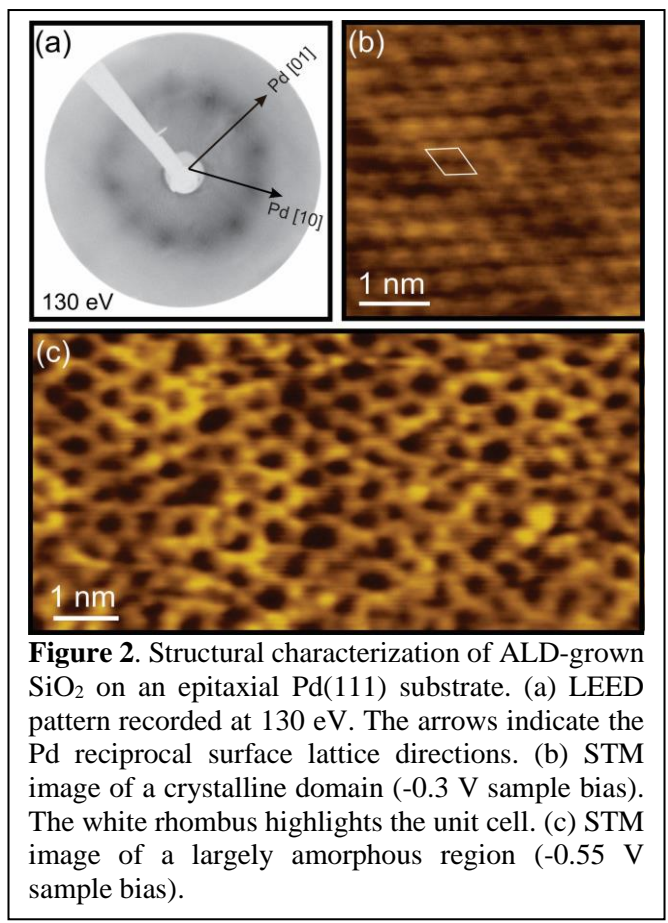

**Figure 2**. Structural characterization of ALD-grown SiO$_2$ on an epitaxial Pd(111) substrate. (a) LEED pattern recorded at 130 eV. The arrows indicate the Pd reciprocal surface lattice directions. (b) STM image of a crystalline domain (-0.3 V sample bias). The white rhombus highlights the unit cell. (c) STM image of a largely amorphous region (-0.55 V sample bias).

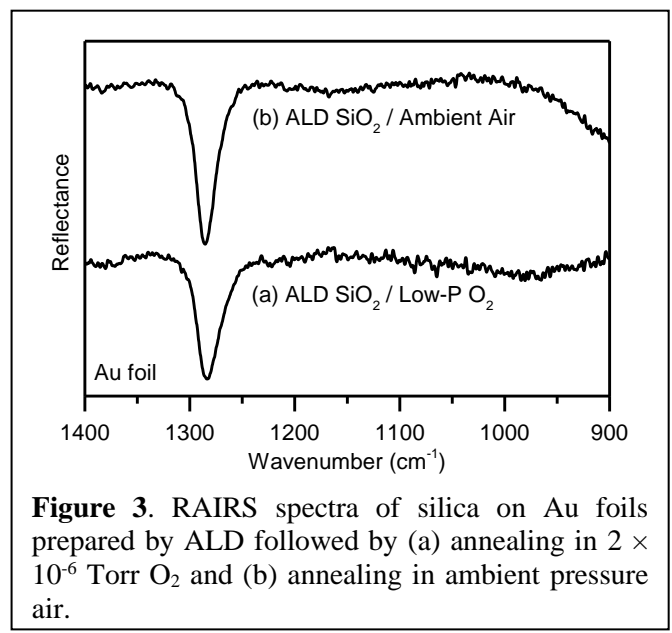

**Figure 3**. RAIRS spectra of silica on Au foils prepared by ALD followed by (a) annealing in $2 \times 10^{-6}$ Torr O$_2$ and (b) annealing in ambient pressure air.

To test whether the ALD process for 2D-VDW SiO$_2$ could be extended to other metal substrates, the same set of experiments was run for polycrystalline Au foils. Gold was chosen for these experiments because its inertness limits the need for pre-processing and allows the annealing step to be carried out in ambient pressure air. RAIRS spectra for two ALD-grown silica layers on Au are provided in **Figure 3**. The spectrum in **Figure 3a** is for two ALD cycles followed by annealing at 975 K in $2 \times 10^{-6}$ Torr O$_2$ while the spectrum in **Figure 3b** was obtained following two ALD cycles and annealing in atmospheric pressure flowing air. Both spectra show the characteristic Si-O-Si stretch, in this case at 1284 cm$^{-1}$ and 1285 cm$^{-1}$, respectively. AES and XPS of these samples show only Si, O, and Au.

To determine how the deposition method affects the phase of the 2D material, 2D silica was prepared by both SiO MBE and ALD on epitaxial Au(111) films on mica, which facilitated 2D silica transfer to TEM grids. Both the ALD and MBE-deposited layers were annealed in flowing air at 950 K. As visible from **Figure 4a**, the layer grown by SiO MBE is highly crystalline. Meanwhile, predominantly amorphous structures could be found on the layer grown by ALD as shown in **Figure 4b**. These results are characteristic of many images recorded over a wide area on several samples. The implication is that the Si source can play an important role in guiding the phase of the final layer. This contradicts prior reports that the strength of the substrate–oxygen interaction governed whether the crystalline or amorphous phase was favored [34]. Regardless, the ability to produce the amorphous phase is desirable for 2D dielectric applications because it eliminates grain boundaries that create discontinuities in the dielectric constants and trap states within the bandgap [49, 50]. The flexibility of the amorphous structure can also be beneficial in fully covering nanostructured surfaces without gross defects, e.g. grain boundaries.

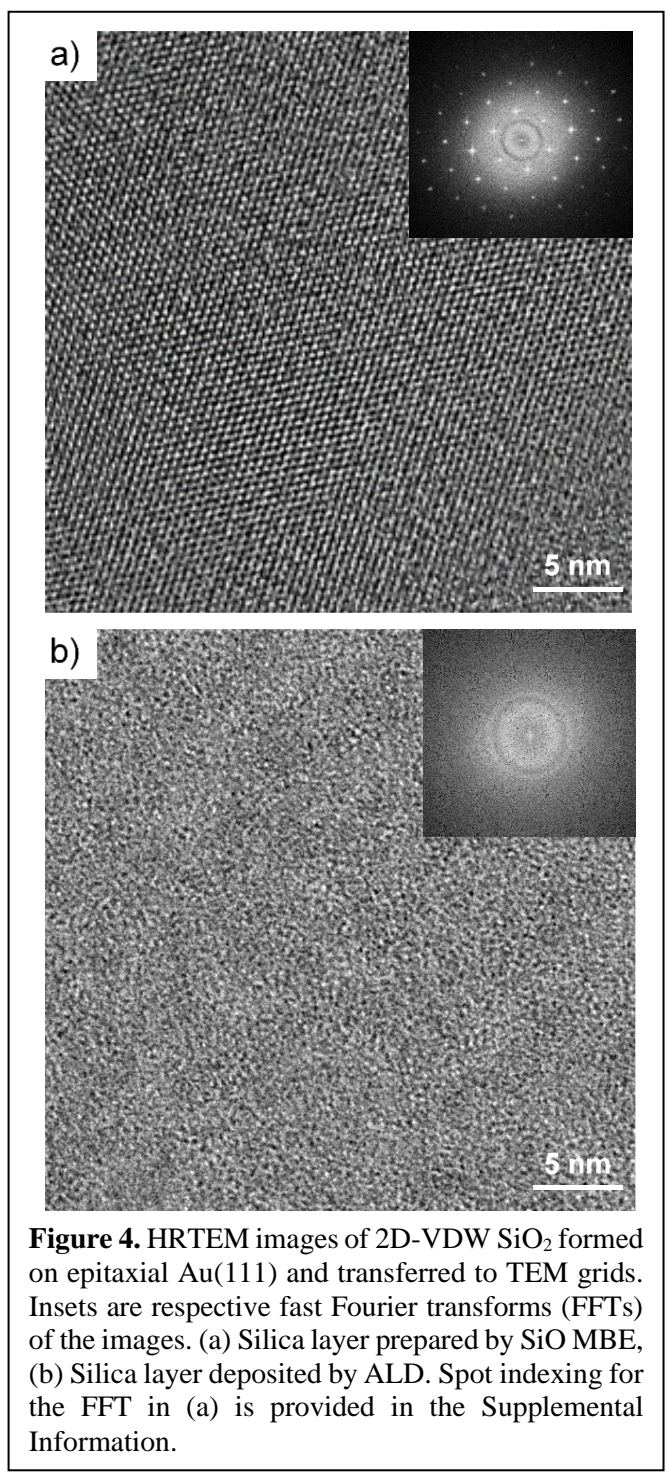

**Figure 4.** HRTEM images of 2D-VDW SiO$_2$ formed on epitaxial Au(111) and transferred to TEM grids. Insets are respective fast Fourier transforms (FFTs) of the images. (a) Silica layer prepared by SiO MBE, (b) Silica layer deposited by ALD. Spot indexing for the FFT in (a) is provided in the Supplemental Information.

To determine the number of ALD cycles required to cover the surface with a silica bilayer, a set of silica samples was prepared by exposing UHV-cleaned Au(111)/mica to 2 – 5 ALD cycles at 525 K followed by annealing in flowing air at 925 K. The samples were then analyzed by RAIRS and AES. As shown in **Figure 5a**, after 2 cycles the characteristic bilayer stretch is seen at 1280 cm$^{-1}$, although with a higher background between 1000 – 1100 cm$^{-1}$ than the other samples. Notably, by 3 cycles the peak in the bilayer region shifts downward toward the range seen for ordered quartz-like structures and peaks emerge between 1100 and 1200 cm$^{-1}$ that are consistent with bulk silica phases [23, 51]. For four and five cycles the features between 1100 and 1200 cm$^{-1}$ continue to dominate. The AES data in **Figure 5b** paint a similar picture. The ratio of the integral of the major Au peak at 64 eV to the O peak at 503 eV for two ALD cycles is comparable to that for 2 ML SiO$_2$/Au(111) single crystal deposited by SiO MBE and annealed in $2\times10^{-6}$ Torr O$_2$, and the Au peak rapidly decays as the number of cycles increased (see Supplemental Information for full AES spectra and further details). These data indicate that for Au at 525 K, two ALD cycles is sufficient to closely reproduce the silica coverage obtained through silica bilayer preparation by SiO MBE and UHV annealing.

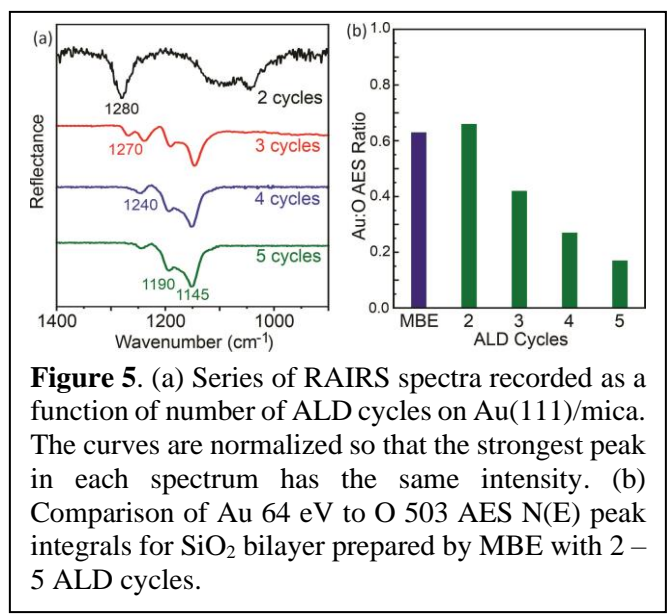

**Figure 5.** (a) Series of RAIRS spectra recorded as a function of number of ALD cycles on Au(111)/mica. The curves are normalized so that the strongest peak in each spectrum has the same intensity. (b) Comparison of Au 64 eV to O 503 AES N(E) peak integrals for SiO$_2$ bilayer prepared by MBE with 2 – 5 ALD cycles.

The same growth procedure was also attempted on Cu foils as an inexpensive substrate. Unfortunately, AES spectra recorded after deposition revealed substantial Cu oxidation, presumably due to the harsh oxygen plasma environment. This oxidation could not be reversed by annealing in vacuum and in the end there was no evidence of 2D silica formation. Given prior reports of 2D silica forming on Cu [21], ALD growth with a weaker oxidant may still be feasible.

## Conclusions

With the successful ALD growth of 2D SiO$_2$ on Pd and Au polycrystalline foils and epitaxial films, we have demonstrated a scalable method to produce bilayer SiO$_2$ that does not require expensive and time-consuming

UHV equipment and processes. ALD is sufficient to deposit the SiO$_x$ precursor at the appropriate thickness, while annealing at 1 atm in either an inert or even open-air environment can produce the 2D material. Moreover, we show that ALD at elevated temperature allows for efficient deposition with only the two ALD cycles anticipated from the 2D silica bilayer structure needed to closely reproduce the coverage obtained by precision MBE. We also demonstrate the potential to control the 2D phase that forms through the Si deposition method with ALD strongly favoring the amorphous phase on Au and SiO MBE producing the crystalline phase. The ALD method unlocks the promise of roll-to-roll production to more easily integrate bilayer SiO$_2$ into VDW heterostructures, and also provides a roadmap to forming single layer 2D material coatings on rough and nanostructured surfaces including catalytic nanoparticles where the confined space between the 2D layer and the catalyst offers exciting opportunities to control catalytic reactions [39-41, 52].

**Methods**

**Silica growth.** Metal foils were purchased from VWR International (Pd: >99.9% metals basis, 0.1 mm thickness; Au: 99.95% metals basis, 0.05 mm thickness). X-ray diffraction of the Pd foil samples revealed predominant (200), (311), and (220) reflections; the intensity of the (111) reflection was smaller by an order of magnitude, which is the only facet with a near match to bilayer SiO$_2$ (0.275 nm hexagonal surface lattice constant vs. 0.265 nm for half the repeat distance of freestanding crystalline bilayer SiO$_2$). To facilitate RAIRS measurements, the foils were mechanically polished to a mirror finish with a progressive sequence of alumina polishing powders. Pd foils were further cleaned with Ar$^+$ sputtering and annealing in UHV and checked with Auger electron spectroscopy (AES) to confirm the absence of residual impurities.

Epitaxial Pd(111)/Cr$_2$O$_3$(0001)/α-Al$_2$O$_3$(0001) films with a Pd thickness of 50 nm were prepared as described previously.[53] Epitaxial Au on mica substrates were obtained from either Phasis Sàrl or Georg Albert PVD. Results for the two sources of epitaxial Au substrates were indistinguishable. Prior to silica deposition the Au on mica samples were annealed in flowing air at 875 K; for the MBE growth the thin film surfaces were also cleaned by sputter/anneal cycles prior to SiO deposition.

ALD deposition was performed using a Veeco Fiji G2 Plasma Enhanced ALD system. To avoid impurities during growth, a two-step pre-conditioning procedure was carried out prior to deposition. The reactor chamber was first baked at 575 K, with the addition of 40 water pulse (0.2 sec per pulse) to neutralize possible residuals in the reactor. Then the chamber was cooled to 525 K, and 40 SiO$_2$ "dummy cycles" were performed to set the desired SiO$_2$ growth environment. After the preconditioning steps, the sample was loaded and then treated by the oxygen plasma at 300 watts plasma power for 2 min to remove residual C accumulated during atmospheric exposure. Finally, the formal ALD growth of SiO$_2$ was achieved. Each ALD cycle started with a 0.06 sec pulse of the silane precursor, followed by a 15 sec pause. Then the O$_2$ flow rate was raised to 50 sccm (standard cubic cm per minute), followed by a 20 sec O$_2$ plasma pulse (300 watts), and ended with another 10 sec pause. The sample temperature was fixed at 525 K while the reactor pressure was kept near $4 \times 10^{-4}$ Torr.

MBE growth was performed by depositing SiO with the sample held at room temperature in $2 \times 10^{-6}$ Torr O$_2$. A high temperature effusion cell (DCA Instruments) was used to sublime SiO. The deposition was monitored with a quartz crystal microbalance to achieve a silicon coverage commensurate with the density of crystalline 2D VDW silica, i.e. $1.64 \times 10^{15}$ Si atoms/cm$^2$.

Final annealing of the samples in UHV was conducted under $2 \times 10^{-6}$ Torr O$_2$ with the temperature measured by an external pyrometer. Ambient pressure annealing was performed under either pure N$_2$ or air in a tube furnace.

**Characterization.** All polarization modulation RAIRS experiments were recorded on a ThermoFisher Nicolet iS50 FTIR spectrometer at a grazing angle of 82°. Baseline correction was applied consistently across samples to minimize introduction of artifacts. Samples were kept in air prior to measurement.

STM images and LEED patterns were recorded in a UHV chamber with a base pressure of $1 \times 10^{-10}$ Torr, as described previously.[24] Cut-and-pulled PtIr alloy tips biased were used for imaging the SiO$_2$ bilayer on Pd/Cr$_2$O$_3$/α-Al$_2$O$_3$. The tunneling current setpoint for the constant current images was set between 0.1 and 1.0 nA; over this range the images were not sensitive to the tunneling current. AES data were collected in the same system using a double-pass cylindrical mirror analyzer at a primary beam energy of 3 keV; the data were collected as N(E) vs. Energy and then numerically differentiated to obtain standard dN(E)/dE vs. Energy spectra. XPS measurements were performed using a separate PHI VersaProbe II system.

HRTEM images were recorded in a field emission transmission electron microscope JEOL JEM 2200FS operating at 200 keV with the CMOS camera "OneView" from Gatan. The Au/mica substrates carrying 2D silica were first coated with poly(methyl methacrylate) (PMMA) and then etched with $I_2$/KI in water. Following the rinse with ultrapure water, 2D silica was transferred onto Quantifoil Multi A TEM grids (Quantifoil Micro Tools GmbH), and the PMMA layers were dissolved in acetone. The contrast between the 2D silica and the TEM grid was obvious (see Supplemental Information).

**Acknowledgements**

**Funding**

**References**

[1]  Geim A K and Grigorieva I V 2013 Van der Waals Heterostructures *Nature* **499** 419
[2]  Novoselov K S, Mishchenko A, Carvalho A and Castro Neto A H 2016 2D Materials and van der Waals Heterostructures *Science* **353** 6298
[3]  Zeng M, Xiao Y, Liu J, Yang K and Fu L 2018 Exploring Two-Dimensional Materials toward the Next-Generation Circuits: From Monomer Design to Assembly Control *Chemical Reviews* **118** 6236-96
[4]  Chaves A, Azadani J G, Alsalman H, da Costa D R, Frisenda R, Chaves A J, Song S H, Kim Y D, He D, Zhou J, Castellanos-Gomez A, Peeters F M, Liu Z, Hinkle C L, Oh S-H, Ye P D, Koester S J, Lee Y H, Avouris P, Wang X and Low T 2020 Bandgap engineering of two-dimensional semiconductor materials *npj 2D Materials and Applications* **4** 29
[5]  Bhaviripudi S, Jia X, Dresselhaus M S and Kong J 2010 Role of Kinetic Factors in Chemical Vapor Deposition Synthesis of Uniform Large Area Graphene Using Copper Catalyst *Nano Letters* **10** 4128-33
[6]  Fan Y, Li L, Yu G, Geng D, Zhang X and Hu W 2021 Recent Advances in Growth of Large-Sized 2D Single Crystals on Cu Substrates *Advanced Materials* **33** 2003956
[7]  Cai Z, Liu B, Zou X and Cheng H-M 2018 Chemical Vapor Deposition Growth and Applications of Two-Dimensional Materials and Their Heterostructures *Chemical Reviews* **118** 6091-133
[8]  Wang D, Luo F, Lu M, Xie X, Huang L and Huang W 2019 Chemical Vapor Transport Reactions for Synthesizing Layered Materials and Their 2D Counterparts *Small* **15** 1804404
[9]  Stoica T, Stoica M, Duchamp M, Tiedemann A, Mantl S, Grützmacher D, Buca D and Kardynał B E 2016 Vapor transport growth of MoS2 nucleated on SiO2 patterns and graphene flakes *Nano Research* **9** 3504-14
[10] Tsakonas C, Dimitropoulos M, Manikas A C and Galiotis C 2021 Growth and in situ characterization of 2D materials by chemical vapour deposition on liquid metal catalysts: a review *Nanoscale* **13** 3346-73
[11] Sonde S, Dolocan A, Lu N, Corbet C, Kim M J, Tutuc E, Banerjee S K and Colombo L 2017 Ultrathin, wafer-scale hexagonal boron nitride on dielectric surfaces by diffusion and segregation mechanism *2D Materials* **4** 025052
[12] Zhou C, Liang X, Hutchings G S, Fishman Z S, Jhang J-H, Schwarz U D, Ismail-Beigi S and Altman E I 2019 Stucture of a Two-dimensional Silicate Formed by Reaction with an Alloy Substrate *Chemistry of Materials* **31** 851
[13] Dong J, Zhang L, Dai X and Ding F 2020 The epitaxy of 2D materials growth *Nature Communications* **11** 5862
[14] Walsh L A and Hinkle C L 2017 van der Waals epitaxy: 2D materials and topological insulators *Applied Materials Today* **9** 504-15
[15] Chiappe D, Ludwig J, Leonhardt A, El Kazzi S, Nalin Mehta A, Nuytten T, Celano U, Sutar S, Pourtois G, Caymax M, Paredis K, Vandervorst W, Lin D, De Gendt S, Barla K, Huyghebaert C, Asselberghs I and Radu I 2018 Layer-controlled epitaxy of 2D semiconductors: bridging nanoscale phenomena to wafer-scale uniformity *Nanotechnology* **29** 425602
[16] Zhang Y, Ren W, Jiang Z, Yang S, Jing W, Shi P, Wu X and Ye Z-G 2014 Low-temperature remote plasma-enhanced atomic layer deposition of graphene and characterization of its atomic-level structure *Journal of Materials Chemistry C* **2** 7570-4
[17] Vervuurt R H J, Kessels W M M and Bol A A 2017 Atomic Layer Deposition for Graphene Device Integration *Advanced Materials Interfaces* **4** 1700232
[18] Cai J, Han X, Wang X and Meng X 2020 Atomic Layer Deposition of Two-Dimensional Layered Materials: Processes, Growth Mechanisms, and Characteristics *Matter* **2** 587-630
[19] Büchner C and Heyde M 2017 Two-dimensional silica opens new perspectives *Progress in Surface Science* **92** 341-74
[20] Heyde M, Shaikhutdinov S and Freund H J 2012 Two-dimensional Silica: Crystalline and Vitreous *Chemical Physics Letters* **550** 1-7
[21] Huang P Y, Kurasch S, Srivastava A, Skakalova V, Kotakoski J, Krasheninnikov A V, Hovden R, Mao Q, Meyer J C, Smet J, Muller D A and Kaiser U 2012 Direct Imaging of a Two-Dimensional Silica Glass on Graphene *Nano Letters* **12** 1081-6
[22] Lichtenstein L, Buchner C, Yang B, Shaikhutdinov S, Heyde M, Sierka M, Wlodarczyk R, Sauer J and Freund H J 2012 The Atomic Structure of a Metal-

[23] Yang B, Kaden W E, Xin Y, Boscoboinik J A, Martynova Y, Lichtenstein L, Heyde M, Sterrer M, Wlodarczyk R, Sierka M, Sauer J, Shaikhutdinov S and Freund H J 2012 Thin Silica Films on Ru(0001): Monolayer, Bilayer and Three-dimensional Networks of [SiO$_4$] Tetrahedra *Physical Chemistry Chemical Physics* **14** 11344-51

supported Vitreous Thin Silica Film *Angewandte Chemie International Edition* **51** 404-7

[24] Altman E I, Götzen J, Samudrala N and Schwarz U D 2013 Growth and Characterization of Crystalline Silica Films on Pd(100) *Journal of Physical Chemistry C* **117** 26144-55

[25] Altman E I and Schwarz U D 2014 Structural and Electronic Heterogeneity of Two Dimensional Amorphous Silica Layers *Advanced Materials Interfaces* **1** 1400108

[26] Altman E I 2017 Group III Phosphates as Two-Dimensional van der Waals Materials *The Journal of Physical Chemistry C* **121** 16328-41

[27] Zhou C, Liang X, Hutchings G S, Jhang J-H, Schwarz U D, Ismail-Beigi S and Altman E I 2019 Tuning Two-dimensional Phase Formation Through Epitaxial Strain and Growth Conditions: Silica and Silicate on Ni$_x$Pd$_{1-x}$(111) Alloy Substrates *Nanoscale* DOI:10.1039/c9nr05944j

[28] Boscoboinik J A, Yu X, Yang B, Fischer F D, Wodarczyk R, Sierka M, Shaikhutdinov S, Sauer J and Freund H-J 2012 Modeling Zeolites with Metal-supported Two-dimensional Aluminosilicate Films *Angewandte Chemie, International Edition* **51** 6005-8

[29] Włodarczyk R, Sauer J, Yu X, Boscoboinik J A, Yang B, Shaikhutdinov S and Freund H-J 2013 Atomic Structure of an Ultrathin Fe-Silicate Film Grown on a Metal: A Monolayer of Clay? *Journal of the American Chemical Society* **135** 19222-8

[30] Brigatti M F, Galán E and Theng B K G 2013 *Developments in Clay Science,* ed F Bergaya and G Lagaly: Elsevier) pp 21-81

[31] Fischer F D, Sauer J, Yu X, Boscoboinik J A, Shaikhutdinov S and Freund H-J 2015 Ultrathin Ti-Silicate Film on a Ru(0001) Surface *The Journal of Physical Chemistry C* **119** 15443-8

[32] Yu X, Yang B, Boscoboinik J A, Shaikhutdinov S and Freund H-J 2012 Support effects on the atomic structure of ultrathin silica films on metals *Applied Physics Letters* **100** 151608

[33] Zhou C, Liang X, Hutchings G S, Jhang J-H, Schwarz U D, Ismail-Beigi S and Altman E I 2019 Tuning Two-dimensional Phase Formation Through Epitaxial Strain and Growth Conditions: Silica and Silicate on Ni$_x$Pd$_{1-x}$(111) Alloy Substrates *Nanoscale* **11** 21340

[34] Lewandowski A L, Tosoni S, Gura L, Yang Z, Fuhrich A, Prieto M J, Schmidt T, Usvyat D, Schneider W-D, Heyde M, Pacchioni G and Freund H-J 2021 Growth and Atomic-Scale Characterization of Ultrathin Silica and Germania Films: The Crucial Role of the Metal Support *Chemistry* **27** 1870-85

[35] Klemm H W, Peschel G, Madej E, Fuhrich A, Timm M, Menzel D, Schmidt T and Freund H J 2016 Preparation of silica films on Ru(0001): A LEEM/PEEM study *Surface Science* **643** 45-51

[36] Yao B, Mandrà S, Curry J O, Shaikhutdinov S, Freund H-J and Schrier J 2017 Gas Separation through Bilayer Silica, the Thinnest Possible Silica Membrane *ACS Applied Materials & Interfaces* **9** 43061-71

[37] Hutchings G S and Altman E I 2019 An atomically thin molecular aperture: two-dimensional gallium phosphate *Nanoscale Horizons* **4** 667-73

[38] Boscoboinik J A and Shaikhutdinov S 2014 Exploring Zeolite Chemistry with the Tools of Surface Science: Challenges, Opportunities, and Limitations *Catal Lett* 1-9

[39] Jhang J-H, Boscoboinik J A and Altman E I 2020 Ambient pressure x-ray photoelectron spectroscopy study of water formation and adsorption under two-dimensional silica and aluminosilicate layers on Pd(111) *The Journal of Chemical Physics* **152** 084705

[40] Prieto M J, Mullan T, Schlutow M, Gottlob D M, Tănase L C, Menzel D, Sauer J, Usvyat D, Schmidt T and Freund H-J 2021 Insights into Reaction Kinetics in Confined Space: Real Time Observation of Water Formation under a Silica Cover *Journal of the American Chemical Society* **143** 8780-90

[41] Eads C N, Boscoboinik J A, Head A R, Hunt A, Waluyo I, Stacchiola D J and Tenney S A 2021 Enhanced Catalysis under 2D Silica: A CO Oxidation Study *Angewandte Chemie International Edition* **60** 10888-94

[42] Büchner C, Wang Z-J, Burson K M, Willinger M-G, Heyde M, Schlögl R and Freund H-J 2016 A Large-Area Transferable Wide Band Gap 2D Silicon Dioxide Layer *ACS Nano* **10** 7982

[43] Jhang J-H, Zhou C, Dagdeviren O E, Hutchings G S, Schwarz U D and Altman E I 2017 Growth of two dimensional silica and aluminosilicate bilayers on Pd(111): from incommensurate to commensurate crystalline *Physical Chemistry Chemical Physics* **19** 14001-11

[44] Tissot H, Weng X, Schlexer P, Pacchioni G, Shaikhutdinov S and Freund H-J 2018 Ultrathin silica films on Pd(111): Structure and adsorption properties *Surface Science* **678** 118-23

[45] Dingemans G, van Helvoirt C A A, Pierreux D, Keuning W and Kessels W M M 2012 Plasma-Assisted ALD for the Conformal Deposition of SiO2: Process, Material and Electronic Properties *Journal of The Electrochemical Society* **159** H277-H85

[46] Włodarczyk R, Sierka M, Sauer J, Löffler D, Uhlrich J, Yu X, Yang B, Groot I, Shaikhutdinov S and Freund H J 2012 Tuning the Electronic Structure of

Ultrathin Crystalline Silica films on Ru(0001) *Physical Review B* **85** 085403

[47] Jhang J-H and Altman E I 2019 Water chemistry on two-dimensional silicates studied by density functional theory and temperature-programmed desorption *Surface Science* **679** 99-109

[48] Klemm H W, Prieto M J, Xiong F, Hassine G B, Heyde M, Menzel D, Sierka M, Schmidt T and Freund H-J 2020 A Silica Bilayer Supported on Ru(0001): Following the Crystalline-to Vitreous Transformation in Real Time with Spectro-microscopy *Angewandte Chemie International Edition* **59** 10587-93

[49] Manchanda L, Busch B, Green M L, Morris M, Dover R B v, Kwo R and Aravamudhan S 2001 High K gate dielectrics for the silicon industry. In: *Extended Abstracts of International Workshop on Gate Insulator. IWGI 2001 (IEEE Cat. No.01EX537),* pp 56-60

[50] McKenna K P and Shluger A L 2011 Electron and hole trapping in polycrystalline metal oxide materials *Proceedings of the Royal Society A: Mathematical, Physical and Engineering Sciences* **467** 2043-53

[51] Koike C, Noguchi R, Chihara H, Suto H, Ohtaka O, Imai Y, Matsumoto T and Tsuchiyama A 2013 Infrared Spectra of Silica Polymorphs and the Conditions of their Formation *The Astrophysical Journal* **778** 60

[52] Shifa T A and Vomiero A 2019 Confined Catalysis: Progress and Prospects in Energy Conversion *Advanced Energy Materials* **9** 1902307

[53] Hutchings G S, Jhang J-H, Zhou C, Hynek D, Schwarz U D and Altman E I 2017 Epitaxial $Ni_xPd_{1-x}$ (111) Alloy Substrates with Continuously Tunable Lattice Constants for 2D Materials Growth *ACS Applied Materials & Interfaces* **9** 11266-71

**Supporting Information**

**Scalable production of single 2D van der Waals layers through atomic layer deposition: Bilayer silica on metal foils and films**

**SI 1. Detecting Excess Silica**

The reflection absorption infrared spectroscopy (RAIRS) can sensitively distinguish silicon deposition beyond that required to form the two-dimensional van der Waals (2D-VDW) bilayer. As detailed in the main text, the 180° Si–O–Si bond creates a distinctive RAIRS peak near 1290 cm$^{-1}$, in contrast to oriented quartz layers which show peaks near 1265 cm$^{-1}$ [1]. As illustrated in Fig. SI 1, exceeding the target Si coverage by just 0.3 ML severely attenuates the bilayer peak near 1290 cm$^{-1}$ and creates an easily detectable sharp peak at 1265 cm$^{-1}$ assigned to ordered quartz structures.

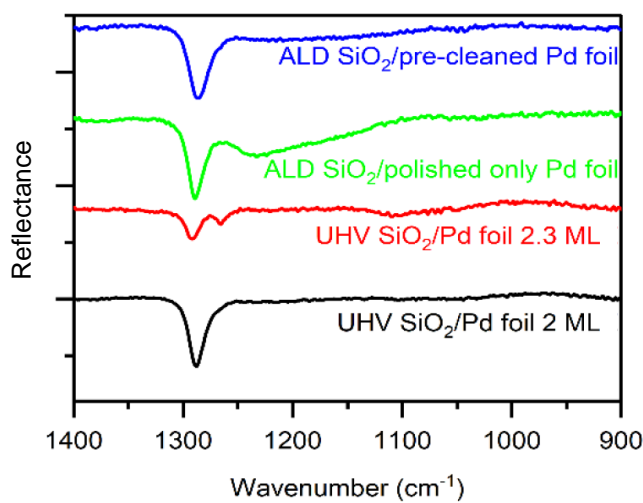

**Figure SI 1.** Comparison of RAIRS spectra for different silica film preparations on polycrystalline Pd foil. UHV SiO$_2$ refers to films prepared by SiO deposition followed by annealing, both in UHV. Black and red curves correspond to 2.0 and 2.3 ML SiO, respectively. The green curve corresponds to an ALD deposited and annealed layer on a polished foil that was not further processed before deposition, while the blue curve corresponds to a polished foil that was cleaned by sputter-anneal cycles prior to ALD silica deposition and annealing.

**SI.2. LEED Analysis of ALD Silica on Pd(111)**

LEED data was collected at electron energies between 45 – 200 eV for ALD SiO$_2$ deposited onto epitaxial Pd(111) which provides a more comprehensive view of all the features seen compared to a pattern collected at a single energy. Representative patterns from this energy range are provided in Fig. SI2. Note the clear spots that emerge at 55 and 85 eV, and the closely spaced but distinct spots that within the cyan circle at 85 eV. Also note that the continuous ring is clear at 55 eV, difficult to see at 75 and 85 eV and reemerges at 120 eV. Figure SI3 illustrates all the spots that are seen, which can be explained by two hexagonal domains of 2D crystalline SiO$_2$ with a lattice constant of 5.3 Å, with one domain aligned with the Pd substrate close-packed direction and the other rotated by 30° [2].

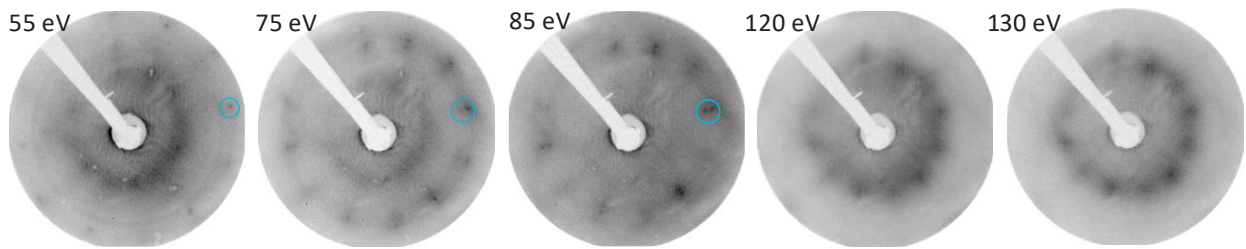

**Figure SI 2.** Series of LEED patterns collected at different energies for 2D silica deposited onto epitaxial Pd(111) by two ALD cycles followed by annealing in 2×10$^{-6}$ Torr O$_2$. The cyan circles highlight a location where a single spot is seen at 55 eV and a second spot emerges at 75 eV and becomes equally intense as the original spot at 85 eV.

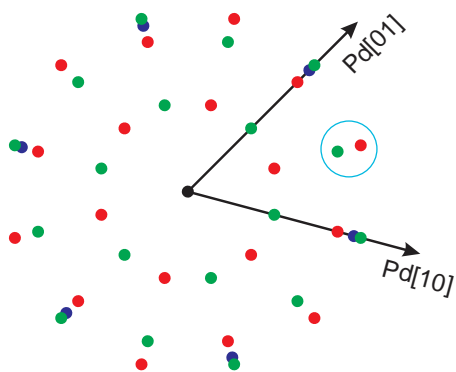

**Figure SI 3.** Model of the spot positions seen in the LEED patterns in Fig. SI2. The partially obscured blue dots represent the positions of the diffraction spots for the bare Pd(111) surface. The green and red dots represent two hexagonal domains, one aligned with the Pd surface [10] direction, the other rotated by 30°. The cyan circle highlights the region where two closely spaced spots are seen in the experimental LEED pattern.

### SI.3 XPS Analysis of ALD Silica Layers Annealed in N$_2$

To determine if the amine group in the precursor of annealing the ALD-deposited silica layer on Pd in atmospheric pressure N$_2$ introduces N impurities into the 2D-VDW silica layer deposited by ALD, samples were analyzed by x-ray photoelectron spectroscopy (XPS). The data in Fig. SI 2 show the expected Si:O ratio for SiO$_2$, that the nitrogen is below the detection limit when the samples are annealed in the UHV system and that a nitrogen impurity just above the detection limit is seen when the samples are annealed in atmospheric pressure N$_2$.

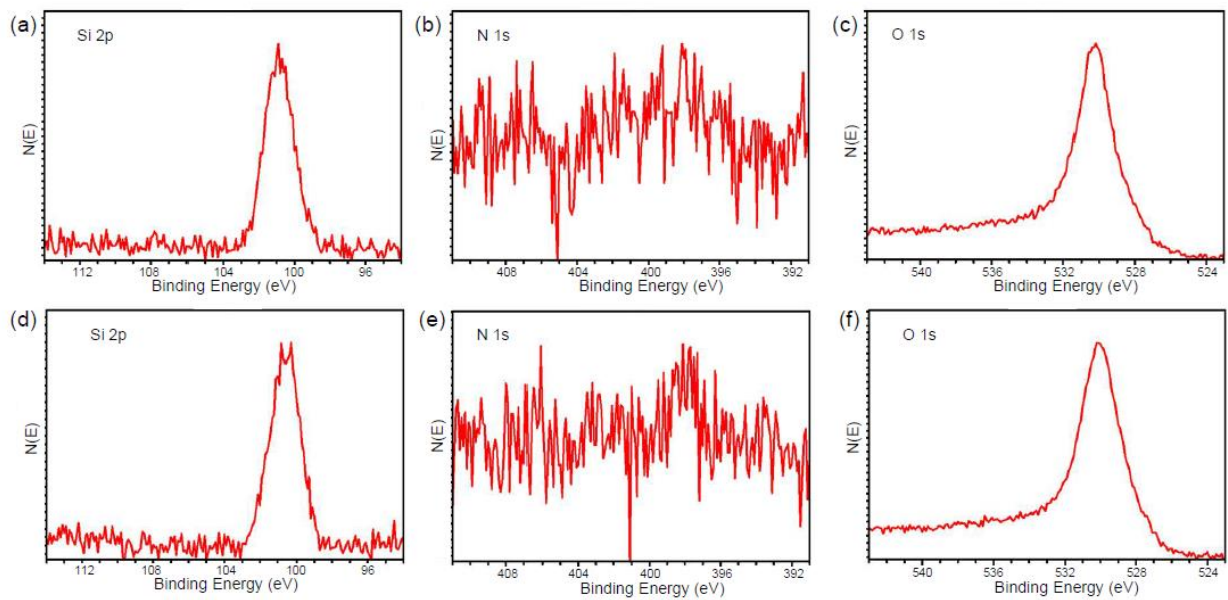

**Figure SI 4.** a - c) XPS core level spectra recorded for an ALD-deposited silica layer on Pd(111) annealed in low pressure O$_2$ in the UHV system. a) Si 2p, b) N 1s, and c) O 1s core levels. d – f) Same as a – c) except the ALD-deposited silica layer was annealed in atmospheric pressure flowing N$_2$. d) Si 2p, e) N 1s, and f) O 1s core levels. The O1s core level overlaps the Pd 3p$_{3/2}$ peak; the assymetry at higher energy is due to Pd 3p$_{3/2}$ emission.

## SI.4 Analysis of HRTEM FFT

The FFT pattern in the inset in Fig. 4a is reproduced below in Fig. SI5. The red rhombus indicates the 2D reciprocal unit cell, showing that the spots are due to multiple diffraction orders of the primitive reciprocal unit cell of crystalline 2D silica in plan view. The labeling indicates the spot indecis for selected spots out to third order spots.

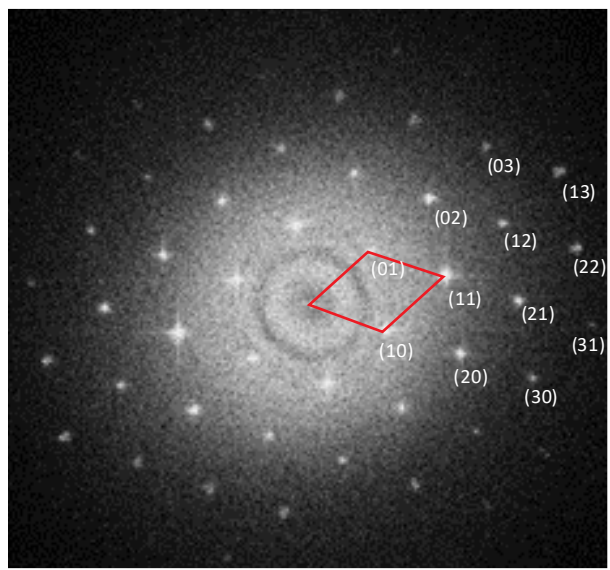

**Figure SI 5.** Inset of Fig. 4a., FFT of 2D silica crystalline domain formed by MBE growth on Au(111)/mica. The red lines highlight the primitive 2D reciprocal unit cell. The white labels are the indecis just up and to the left of the labels.

## SI.5 AES Analysis of 2D Silica

Auger electron spectroscopy was used to charcterize the silica coverage obtained by ALD deposition and to monitor impurities. The data were collected as N(E) vs Energy and then numerically differentiated to obtain the spectra in the typical dN(E)/dE versus energy form. Figure SI6 shows a series of spectra obtained for 2 – 5 ALD cycles on Au(111)/mica followed by annealing in flowing air; for comparison the figure also shows spectra for a Au(111) single crystal obtained following deposition of 2 ML SiO by MBE and annealing in $2 \times 10^{-6}$ Torr $O_2$. The ALD samples were not treated in any way between insertion into UHV and collecting the spectra. The most intense Au and Si peaks are seen between 40 – 100 eV in Fig. SI6a while Fig. SI6b shows the primary oxygen peak at 503 eV a series of weaker Au peaks between 140 – 240 eV and covers the range where impurities are typically seen. These data in the N(E) vs. Energy form were used to obtain the peak integral bar graph in Fig. 5b of the main text. Several points emerge from the spectra. First, several impurities can be detected on the ALD samples, C, S, and K, that are not seen on the sample entirely prepared in the UHV system. We find for samples prepared in vacuum, exposed to the ambient and returned to vacuum that such impurities can be removed by annealing in low pressure $O_2$ at 600 – 900 K. More significantly, note that the Au peaks are severely attenuated by three ALD cycles, are difficult to see at four cycles, and appear completely obscured by five cycles. This is borne out in Fig. 5b of the main text which shows that the Au (64 eV):O (503 eV) peak integral ratio for two cycles is comparable to that seen for the MBE sample and then rapidly decays thereafter. Note that the O peak is used as a proxy for the $SiO_2$ coverage because of the overlap of the Si AES peak with the tail of the 64 eV Au peak; since $O_2$ does not adsorb or react with Au, all of the oxygen can be assumed to be associated with Si.

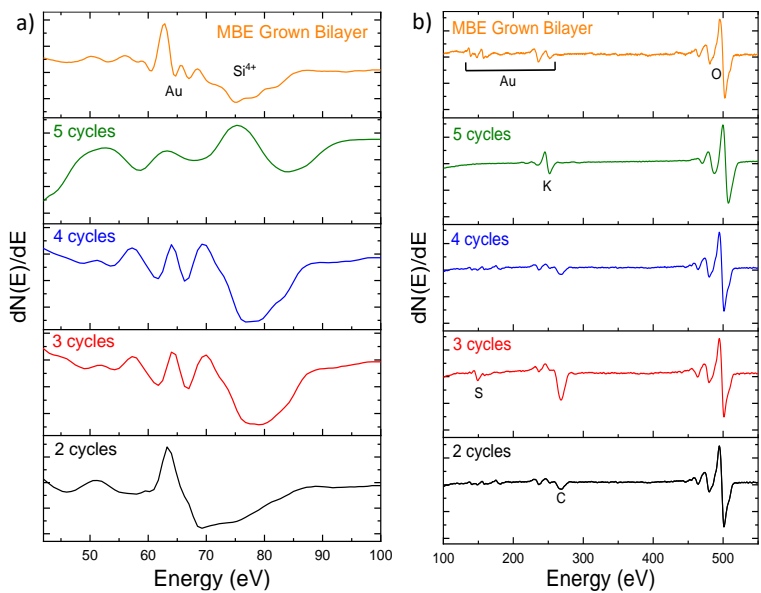

**Figure SI 6.** AES spectra covering (a) 40 – 100 eV and (b) 100 – 540 eV for a MBE grown $SiO_2$ bilayer on an Au(111) single crystal and series of ALD grown and ambient pressure annealed silica films.

## SI.6 Distinction Between TEM Grid and 2D Silica

It can sometimes be difficult to distinguish atomically thin amorphous materials from an amorphous TEM grid. Figure SI7 shows that there is easily distinguishable contrast between the atomically thin silica layer and the TEM grid. The data in the figure are for the MBE grown 2D silica.

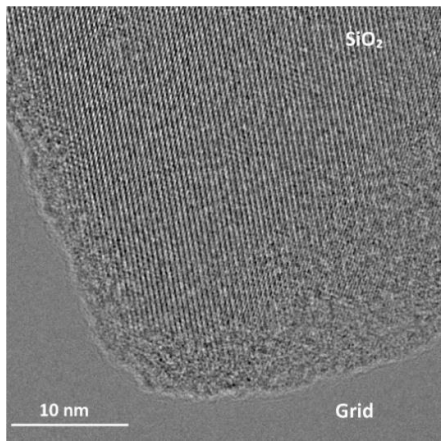

**Figure SI 7.** HRTEM image showing the TEM grid partly covered by crystalline 2D $SiO_2$. The contrast between the grid and the silica allows easy distinction. The image is of a 2D silica layer prepared by SiO deposition in UHV followed by annealing in flowing air at atmospheric pressure.

**Reference**

[1] Yang B, Kaden W E, Xin Y, Boscoboinik J A, Martynova Y, Lichtenstein L, Heyde M, Sterrer M, Wlodarczyk R, Sierka M, Sauer J, Shaikhutdinov S and Freund H J 2012 Thin Silica Films on Ru(0001): Monolayer, Bilayer and Three-dimensional Networks of [SiO4] Tetrahedra *Physical Chemistry Chemical Physics* **14** 11344-51

[2] Jhang J-H, Zhou C, Dagdeviren O E, Hutchings G S, Schwarz U D and Altman E I 2017 Growth of two dimensional silica and aluminosilicate bilayers on Pd(111): from incommensurate to commensurate crystalline *Physical Chemistry Chemical Physics* **19** 14001-11